\documentclass[sigconf,nonacm,dvipsnames]{acmart}

\usepackage{amsfonts}
\usepackage{amsmath}
\usepackage{array}
\usepackage{multirow}
\usepackage{multicol}
\usepackage{todonotes}
\usepackage{lipsum}
\usepackage{tikz}
\usepackage{tcolorbox}
\usepackage{subcaption}

\usetikzlibrary{automata,matrix,arrows.meta,positioning,calc,shapes,shapes.misc}

\tikzset{%
    pics/branch/.style={%
        code={%
            \node[circle,minimum width=1cm] (c0) at (0, 0) {\normalsize $C_{0,1}$};
            \node[circle,minimum width=1cm] (c1) at (2, 0) {\normalsize $C_{0,2}$};
            \node[circle,minimum width=1cm] (c2) at (2, -1) {\normalsize $C_{1,1}$};
            \node[circle,minimum width=1cm] (c3) at (4, -1) {\normalsize $C_{1,2}$};
            \node[circle,minimum width=1cm] (c4) at (8, -1) {\normalsize $C_{1,k}$};
            \node[circle,minimum width=1cm] (c5) at (8, 0) {\normalsize $C_{0,n}$};
            \node[circle,minimum width=1cm] (c6) at (10, 0) {\normalsize $C_{0,n+1}$};
            \node[circle,minimum width=1cm] (c8) at (4, 0) {\normalsize $C_{0,3}$};
            \node[circle,minimum width=1cm] (dots0) at (6, 0) {$\hdots$};
            \node[circle,minimum width=1cm] (dots1) at (6, -1) {$\hdots$};
    
            \draw[->] (c0) to (c1);
            \draw[->] (c1) to (c8);
            \draw[->] (c8) to (dots0);
            \draw[->] (dots0) to (c5);
            \draw[->] (c5) to (c6);
    
            \draw[->] (c0) to (c2);
            \draw[->] (c2) to (c3);
            \draw[->] (c3) to (dots1);
            \draw[->] (dots1) to (c4);
            \draw[->] (c4) to (c6);
        }
    },
    pics/commitpath/.style={%
        code={%
            \node[circle,minimum width=0.75cm] (c11) at (0, 0) {$C_{1, 1}$};
            \node[circle,minimum width=0.75cm] (c12) at (2, 0) {$C_{1, 2}$};
            \node[circle,minimum width=0.75cm] (c13) at (4, 0) {$C_{1, 3}$};
            \node[circle,minimum width=0.75cm] (c14) at (6, 0) {$C_{1, 4}$};
            \node[circle,minimum width=0.75cm] (c15) at (8, 0) {$C_{1, 5}$};
            \node[circle,minimum width=0.75cm] (c16) at (10, 0) {$C_{1, 6}$};
            \node[circle,minimum width=0.75cm] (c21) at (4, -1) {$C_{2, 1}$};
            \node[circle,minimum width=0.75cm] (c22) at (6, -1) {$C_{2, 2}$};

            \draw[->] (c11) to (c12);
            \draw[->] (c12) to (c13);
            \draw[->] (c13) to (c14);
            \draw[->] (c14) to (c15);
            \draw[->] (c15) to (c16);
            
            \draw[->] (c12) to (c21);
            \draw[->] (c21) to (c22);
            \draw[->] (c22) to (c15);
        }
    }
}
\tikzset{cross/.style={cross out, draw=black, minimum size=2*(#1-\pgflinewidth), inner sep=0pt, outer sep=0pt},
cross/.default={1pt}}

\AtBeginDocument{%
  }

\setcopyright{acmlicensed}
\copyrightyear{2026}
\acmYear{2026}
\acmDOI{XXXXXXX.XXXXXXX}
\acmConference[ICSE '26]{International Conference On Software Engineering}{April 12 -- 18,
  2026}{Rio De Janeiro, Brazil}

\begin{document}

\title{Gotta catch 'em all! Towards File Localisation from Issues at Large}

\author{Jesse Maarleveld}
\email{j.maarleveld@rug.nl}
\orcid{0009-0000-7944-1746}
\affiliation{%
  \institution{University of Groningen}
  \city{Groningen}
  \country{The Netherlands}
}

\author{Jiapan Guo}
\email{j.guo@rug.nl}
\orcid{0000-0003-3966-4405}
\affiliation{%
  \institution{University of Groningen}
  \city{Groningen}
  \country{The Netherlands}
}

\author{Daniel Feitosa}
\email{d.feitosa@rug.nl}
\orcid{0000-0001-9371-232X}
\affiliation{%
  \institution{University of Groningen}
  \city{Groningen}
  \country{The Netherlands}
}

\renewcommand{\shortauthors}{Maarleveld et al.}

\begin{abstract}



    Bug localisation, the study of developing methods to localise the files requiring changes to resolve bugs, has been researched for a long time to develop methods capable of saving developers' time. Recently, researchers are starting to consider issues outside of bugs. Nevertheless, most existing research into file localisation from issues focusses on bugs or uses other selection methods to ensure only certain types of issues are considered as part of the focus of the work. Our goal is to work on all issues at large, without any specific selection. 

    In this work, we provide a data pipeline for the creation of issue file localisation datasets, capable of dealing with arbitrary branching and merging practices. We provide a baseline performance evaluation for the file localisation problem using traditional information retrieval approaches. Finally, we use statistical analysis to investigate the influence of biases known in the bug localisation community on our dataset.

    Our results show that methods designed using bug-specific heuristics perform poorly on general issue types, indicating a need for research into general purpose models. Furthermore, we find that there are small, but statistically significant differences in performance between different issue types. Finally, we find that the presence of identifiers have a small effect on performance for most issue types. Many results are project-dependent, encouraging the development of methods which can be tuned to project-specific characteristics. 
\end{abstract}


\maketitle

\section{Introduction}
    Developers spend a significant amount of their time trying to understand code. Research suggests 50\% to 60\% of developer time is spent on code comprehension \cite{ko_exploratory_2006,bohme_where_2017,xia_measuring_2018}. Tools which help developers understand code can thus help save a significant amount of time, improving productivity. This can be especially useful for long-lived systems requiring changes, where tacit knowledge plays an important factor in developer effectiveness~\cite{ryan_acquiring_2013}. One such way of helping developers is by tool-assisted navigation through large and complex code bases while working on specific tasks. Localising the files associated with bugs has been a research interest for a long time \cite{zhou_where_2012}. More recently, preliminary, not yet peer reviewed research is looking into developing automated methods for determining what files need to be changed to resolve issues from issue tracking systems beyond just bugs \cite{chen_locagent_2025,reddy_swerank_2025-1}. 

    Beyond aiding developer in code comprehension, file localisation from general issues can also be used on a larger scale to analyse backlogs. By mapping all open source issues to files that are likely to change, insights can be obtained regarding e.g. the amount of work in the backlog, which in turn can serve other purposes. For example, through combination with SATD detection tooling (e.g. \cite{shivashankar_beacon-td_2025}), generalised file localisation could lead to an issue-centric view of technical debt in a code base -- which could then in turn be combined with other backlog items to estimate the impact of existing technical debt in the system. Furthermore, techniques for file localisation can also be used as parts of other AI based software development tools. For example, LLM-based bug localisation models may require a seed set of files because the full code base does not fit in the token window \cite{jimenez_swe-bench_2024,xia_demystifying_2025}. Techniques for general issue types can serve as a fast and computationally efficient seed set finders for more generalised automated issue resolving models. 



    In this work, we investigate file localisation for issues, without restricting our investigation to any particular type of issue. Most existing research only considers bug localisation (e.g. \cite{zhou_where_2012,xiao_improving_2017,ciborowska_fast_2022,du_pre-training_2023,huo_deep_2021,lam_combining_2015,yang_locating_2021,qin_agentfl_2025,zhou_benchmarking_2025}). Research that considers a more wide variety of issue types frequently still imposes certain constraints like a selection of issue types \cite{chen_locagent_2025} or filtering on test modifications and/or failure \cite{yang_swe-bench_2024,jimenez_swe-bench_2024,zan_multi-swe-bench_2025,reddy_swerank_2025-1}. Moreover, unlike part of existing research (e.g. \cite{reddy_swerank_2025-1}, \cite{chen_locagent_2025}), we do not pose any restrictions on the programming languages that may be used in a project. Finally, our method for dataset creation is independent from an organisation's Git workflow (i.e. branching practices) and issue tracking systems, whereas many modern datasets assume a GitHub pull request-based workflow for data mining \cite{reddy_swerank_2025-1,chen_locagent_2025,yang_swe-bench_2024,jimenez_swe-bench_2024,zan_multi-swe-bench_2025}.
    
    We first created a new data extraction pipeline and dataset which creates a labelled dataset by mining links between issues and commits, using changes files from commits as ground truths. Contrary to the majority of existing bug localisation research, we pay extensive attention to make sure we appropriate handle merge commits and issues with multiple associated commits. Using our mined dataset, we performed an initial exploration of the problem using a number of information retrieval methods. Inspired by research into various biases present in bug localisation dataset which were obtained by similar means as our dataset (e.g. \cite{kochhar_potential_2014,widyasari_influence_2022}), we also investigated a number of factors that may affect performance in our own dataset. In the end, we make the following contributions:

    \begin{itemize}
        \item We provide a data extraction and loading pipeline capable of dealing with complex branching practices and their resulting potential ambiguities.

        \item We provide a ready to use dataset consisting of seven medium sized projects spanning different domains and programming languages. 

        \item We perform an exploratory performance analysis for the file localisation problem for unrestricted issue types using information retrieval methods.

        \item We study how issue type and the number of identifiers in issues affect predictive performance of information retrieval methods.
    \end{itemize}

    The remainder of this paper is organized as follows. In Section \ref{sec:background-related-work}, we will discuss background and related work. In Section \ref{sec:study-design-complete}, we present our study design, including a more formal description of the machine learning problem, the creation of our dataset, and the design of our experiments. We present our results in Section \ref{sec:results} and further discuss them in Section \ref{sec:discussion}, together with the threats to validity. We end with our conclusions and plans for future work in Section \ref{sec:conclusion}.

\section{Background \& Related Work}\label{sec:background-related-work}
    \subsection{Localisation Problems}
        In \textbf{bug localisation} research, the goal is to design models that, given a bug report, predict the files containing said bug or that need to be modified to fix said bug \cite{zhou_where_2012}. In bug localisation, the datasets are acquired by linking commits to bug reports, and then using the files changed in the commit as ground truths \cite{kochhar_potential_2014,kim_are_2021}. Conceptually, the main difference with our research is that we focus on a wider variety of types of issues -- not just bugs.

        Bug localisation has seen a lot of research and different approaches, ranging from more traditional information retrieval methods (e.g. \cite{zhou_where_2012}), to modern deep learning approaches such as convolutional neural networks (e.g. \cite{xiao_improving_2017}), transformer models (e.g. \cite{ciborowska_fast_2022}), graph based neural networks (e.g. \cite{du_pre-training_2023}), deep transfer learning (e.g. \cite{huo_deep_2021}), and various hybrid models which combine multiple information sources, like text in different forms and manually crafted bug related features (e.g. \cite{lam_combining_2015,yang_locating_2021}). Recently, the field is moving towards the use LLM-based AI agents \cite{qin_agentfl_2025,zhou_benchmarking_2025}.

        A number of works from the bug localisation field did go slightly beyond bugs; \citet{canfora_fine_2006} considered change requests in general. They use information retrieval techniques to find old issues similar to new issues, and use version history information for these issue to come up with candidate files for the new issue \cite{canfora_impact_2005,canfora_fine_2006}. This work mostly focuses on bug reports and enhancement/feature requests.

        More recently, early-stage works which have not been peer-reviewed yet, are exploring how LLMs have been used for file localisation. LocAgent \cite{chen_locagent_2025} is an LLM-based AI agent for file localisation. It is meant to be applied on more issue types than just bugs, but still focuses on ``problems'' to be solved (feature requests, performance issues, and security issues). It performs best for bugs, followed by feature requests, security issues, and performance issues \cite{chen_locagent_2025}. SweRank \cite{reddy_swerank_2025-1} uses LLMs for embedding and ranking. It is trained primarily on bug data, but has demonstrated reasonable generalisability towards feature requests, performance issues, and security issues. It outperforms LocAgent, while reducing operational costs. 

        In the mobile app domain, researchers have focussed on \textbf{localising files from app reviews} (e.g. \cite{palomba_recommending_2017,zhang_where2change_2021}). The objective is similar to ours in the sense that the goal is to predict which source files will change based on natural language. However, even though issues are sometimes uses as an auxiliary artefact for these predictions (e.g. \cite{zhang_where2change_2021}), the primary artefacts remains app reviews from users.

    \subsection{Localisation Datasets}

        For bug localisation specifically, a variety of datasets exists. \citet{ye_learning_2014} used a dataset consisting of the projects Aspect4J, Birt, Eclipse Platform UI, JDT, SWT, and Tomcat, where bugs from bug trackers are linked to commits through pattern matching \cite{ye_learning_2014}. This or similar project selections have also been used for evaluation in more recent research (e.g. \cite{zhou_where_2012,ciborowska_fast_2022}). Other datasets include Defects4J \cite{just_defects4j_2014}, Long Code Arena \cite{bogomolov_long_2024}, BuGL \cite{muvva_bugl_2020}, Beetlebox \cite{chakraborty_blaze_2025}, and Bugs.jar \cite{saha_bugsjar_2018}. Common flaws we found in such dataset is that they commonly focus on a single programming language (\citeauthor{ye_learning_2014}, Defects4J, Bugs.Jar), or assume the usage of GitHub issues and Pull Requests for their data pipelines (Long Code Arena, BuGL, BeetleBox).

        SWE-bench \cite{jimenez_swe-bench_2024} is a dataset for automated program repair consisting of around 2,300 issues with fixes. It has spawned a number of other related datasets, such as a multi-lingual variant Multi-SWE-bench \cite{zan_multi-swe-bench_2025}, and a multi-model variant SWE-bench Multimodal \cite{yang_swe-bench_2024}. SWE-bench+ \cite{aleithan_swe-bench_2024} is a variant of SWE-bench that attempts to minimise potential data-leakage. These datasets are meant for the training and validation of models which generate fixes for issues from the given issue description, and consists of issues together with their resolving pull requests and merge commits; Bug localisation is one of the steps in the process. The datasets tend to be biased towards bugs due to the way in which they are constructed \cite{jimenez_swe-bench_2024,chen_locagent_2025}. Only commits that modified a test were included. Furthermore, there must be at least one test instance that failed before applying the change, but passed after \cite{jimenez_swe-bench_2024}.

        Loc-Bench \cite{chen_locagent_2025} is a dataset consisting of 560 issues from 165 different Python repositories. The dataset is meant for the evaluation of models for file localisation for issue types beyond bugs. Specifically, the issue types contained in the dataset are ``Bug Report'', ``Feature Request'', ``Security Issue'', and ``Performance Issue''. SweLoc \cite{reddy_swerank_2025-1} is a dataset consisting of 67,341 issues with associated code from 3387 Python repositories for file localisation. Similar to Swe-bench, commits associated to issues in SweLoc must also modify test files. The dataset is primarily bug-oriented \cite{reddy_swerank_2025-1}.

        \citet{kochhar_potential_2014} identified and evaluated the impact of three potential biases in bug localisation datasets. \citet{widyasari_influence_2022} later repeated their experiments with more data and different bug localisation techniques. The three biases are the following:

        \begin{enumerate}
            \item \textit{Report Misclassification}: Caused by issues which are not bugs, but are labelled as such and thus included in the dataset. The impact of this bias is almost never statistically significant, and if it is, effect size is negligible. 

            \item \textit{Localised Bugs}: Caused by issues which contain identifiers (e.g. in the form of class or file names) of the affected files. The impact of this bias is statistically significant, with partially localised bugs resulting in higher performance scores. 

            \item \textit{Non-buggy files}: Caused by wrongly identified ground truth files, i.e. files which are part of the commit, but do not actually contain the bug. The impact of this bias is not statistically significant.
        \end{enumerate}

\section{Study Design}\label{sec:study-design-complete}
    In this section, we first define the problem more explicitly, and then introduce our research questions. Next, we explained how we created our dataset, what preprocessing and retrieval methods we use, and our chosen evaluation metrics. Finally, we provide an overview of the experiments we performed.

    We note that, compared to existing work, our work does not use any a priori filtering on issue type, nor do we use other selection criteria to select issues with specific characteristics; We consider all issue types at large. We also do not have any restrictions on the programming languages used in the projects in our dataset. Because all this might change the characteristics of the issues in our data compared to more narrowly scoped datasets, we investigate our dataset for the presence of biases previously identified in bug localisation datasets. 

    \subsection{Problem Definition}
        Formally, our goal is the following: given an issue $I$, and a repository branch $B$ containing source files $S_1, \hdots, S_n$, we want to rank all sources files from the most to the least likely to require change to resolve the given issue $I$.

\begin{figure}
            \centering
            \includegraphics[width=\linewidth]{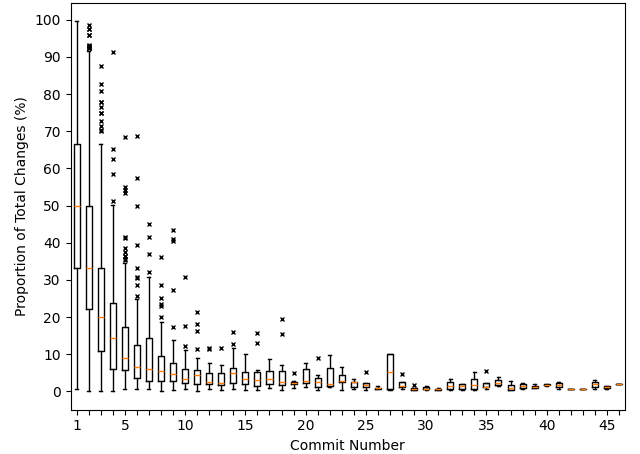}
            \caption{Proportion of the number of changes (as measured by the \# files changed) per commit, measured over all issues with more than one associated commit in our dataset.}
            \label{fig:commit-size}
        \end{figure}

        We use historical issue and commit data from open source projects in order to train and evaluate methods to solve the problem posed above. We use links from commits to issues mined from commit messages as a best-effort proxy to determine the files required to resolve specific issues. Since issues may be resolved over the span of multiple commits, we propose the following set of variations of the problem:
            
        \begin{enumerate}
            \item \textit{First Commit Only}: This version of the problem operates on the assumption that, generally, the first commit related to an issue contains the bulk of the work; Follow up commits contain clean-up or smaller details. As such, focusing on the first commit related to an issue is sufficient. Figure \ref{fig:commit-size} show that the assumption holds reasonably well for our dataset.

            \item \textit{All Future Files}: This version of the problem assumes the work for a single issue might be divided across a number of commits, and that this work is effectively determined up front and is (almost) not influenced by other issues or changes to the system. The idea is that, given the a priori state of the system, the model has to predict all files which will be changed to resolve the issue -- across all related commits. After a commit has been made, the model should then be able to predict all \textit{other remaining} files that still have to change, until the issue is fully resolved.

            \item \textit{Exact Commits}: In this version of the problem, the goal of the model is to predict the changed files for every consecutive related commit, exactly. This requires more feedback signals from changes to the repository (e.g. how other issues or commits reflect the required changes), and also depends on developer intent and/or workflow information. The idea here is that a good prediction model would be able to assist a developer ``every step of the way'' while resolving an issue. 
        \end{enumerate}

        In this study, we focus on the first problem in particular (\textit{First Commit Only}). Our reason is that the second and third formulations of the problem are harder, and therefore not as suitable for an initial exploration of the problem. The latter two variations are harder because they require good feedback signals from other changes, as well as because they might modify files which are not yet present at the time of the first commit. We leave these versions for future work.

    \subsection{Research Questions}
        The goal of this study is to 1) perform an initial exploration of file localisation for issues beyond the scope of bug localisation, 2) provide tooling for dataset creation and loading, and a dataset for file localisation, and 3) understand issue-related factors that may influence the performance of general file localisation models. This is further refined in the following three research questions:
        
        \begin{enumerate}
            \item \textbf{RQ 1: Which information retrieval method performs best for file localisation for general issue types?}

            The earliest models used in bug localisation were information retrieval methods such as TF-IDF \cite{zhou_where_2012}. Even nowadays, simple information retrieval methods can achieve decent performance for text retrieval tasks \cite{abdallah_retrieval_2025}. Hence, such methods can serve as a good performance baseline for more complex future models. 

            \item \textbf{RQ 2: How do issue types affect the performance of file localisation models?}

            Different types of issues tend to be written in different styles. On the one hand, issues for lower level changes like bugs may use words or more similar to or contained in related source files. Some may contain implicit or explicit references to files in the form of file names or identifiers (e.g. function names like \texttt{run\_server} or class names like \texttt{FileReader}) due to the presence of trace backs, or a more detailed planning on how to resolve the issue. On the other hand, issues describing or coordinating more high level  changes to the system may use less specific or more abstract phrasing. The goal of this research question is to investigate to what degree the type of an issue influences the performance of file localisation models. 

            \item \textbf{RQ 3: How does the presence of identifiers and file names affect the performance of file localisation models?}

            Previous research into bug localisation identified the presence of already identified files in issues as a source of performance bias \cite{kochhar_potential_2014,widyasari_influence_2022}. The goal of this research question is to investigate to what degree the more general file localisation problem is affected by this, as measured through the presence of identifiers and file names in issues. We focus on this bias in particular, because 1) in this case, report misclassification does not apply in the same way as it does for bug localisation since we include all issue types, and 2) both report classification and incorrect ground truths were found to have insignificant results in bug localisation \cite{kochhar_potential_2014,widyasari_influence_2022}.  
        \end{enumerate}

    \subsection{Dataset Creation}
        In this section, we describe our process of creating a dataset. First, we explain how we selected the projects used in this study. Next, we explain how we establish links between issues and commits and filter out potential ambiguities. Finally, we explain how to generate the positive and negative samples. 
    
        \subsubsection{Project Selection}
            We aimed to create a dataset containing a variety of different projects. Specifically, we wanted to make sure that we have variety in terms of 1) project domain, 2) organisation, 3) projects scale (both in terms of issues and in terms of size of the code base), 4) project age, and 5) programming languages used. Furthermore, all selected projects must use Git as version control, should report issue IDs in commit messages, and should use a publicly available issue tracking system. In our case, this was Jira; We used the dataset from \cite{maarleveld_maestro_2024}, which is a dataset of Jira issues based on the dataset of \citet{montgomery_alternative_2022}, but extended with additional project information, which we updated with more recent issues\footnote{For all projects in this study, issue data until 13 June 2025 10:00 UTC was fetched.}. We picked a selection of projects with 2,000 to 5,000 issues, excluding any projects where more than 70\% of issues were bugs to ensure issue type diversity. Based on our remaining constraints, we picked a number of projects; For each project we considered selecting, we manually checked the 100 most recent issues to make sure the issue tracking system was used to discuss development (e.g. it was not used for user reports). The final selection is shown in Table \ref{tab:ds-projects}. In the remainder of the text, Spring Data MongoDB will be abbreviate as DMDB.
    
            \begin{table*}[]
                \centering
                \caption{Projects selected to be in our dataset.}
                \label{tab:ds-projects}
                \begin{tabular}{c c c c c}
                    \toprule 
                    Project Name    & Organisation  & Main Programming Languages    & \# Commits    & \# Issues \\ \midrule 
                    Apache Avro & Apache & c, c\#, c++, java, php, python, ruby & 4659 & 4112\\ 
                    Apache Maven & Apache & java & 15546 & 7297\\ 
                    Apache Thrift & Apache & c, c\#, c++, go, java, javascript, ocaml, python, rust, smalltalk, and more & 7146 & 5856\\ 
                    Apache Tika & Apache & java & 9361 & 4393\\ 
                    Apache TomEE & Apache & java & 15584 & 4021\\ 
                    Spring Data MongoDB & Spring & java & 3989 & 2590\\ 
                    Spring Roo & Spring & java & 6404 & 3987 \\
                    \bottomrule 
                \end{tabular}
            \end{table*}
    
        \subsubsection{Commit to Issue Linking}\label{sec:datast-2}
            \paragraph{Raw Link Mining} For each project, we cloned its Git repository\footnote{Version as of 13 June, 2025 10:00 UTC}. We traversed all commit messages and used regex to find links from commits to issues. The regex we used was \texttt{KEY-\textbackslash d+}, where \texttt{KEY} is a project-specific identifier (e.g. \texttt{AVRO} for Apache Avro). We only looked for issue links in the first line of the commit because during manual checking, we found that considering the whole body of the commit message tends to lead to more false positives (e.g. through merge commits summarising the commit messages of a number of other commits). If a commit message mentions multiple issues, we make the assumptions that the mentioned issues involve similar code changes, and that all files modified in the commit would require modification to resolve each individual issue separately.
    
\begin{figure*}
                        \centering
            
            
            
            
            
                        \begin{subfigure}[t]{0.49\textwidth}
                            \centering
                            \scalebox{0.75}{
                                \begin{tikzpicture}
                                    \pic[draw=black] at (0, 0) {commitpath=1};
            
                                    \node[ellipse,draw=OliveGreen,dotted,minimum width=0.75cm,minimum height=0.75cm,line width=2pt] (g0) at (c13) {}; 
                                    \node[ellipse,draw=OliveGreen,dotted,minimum width=0.75cm,minimum height=0.75cm,line width=2pt] (g0) at (c14) {}; 
            
                                    \path[draw=blue,style=->] (0, 0.25) -- (10, 0.25);
                                \end{tikzpicture}
                            }
                            \caption{A path passing through two commits.}
                        \end{subfigure}\quad
                        \begin{subfigure}[t]{0.49\textwidth}
                            \scalebox{0.75}{
                                \begin{tikzpicture}
                                    \pic[draw=black] at (0, 0) {commitpath=1};
            
                                    \node[ellipse,draw=OliveGreen,dotted,minimum width=0.75cm,minimum height=0.75cm,line width=2pt] (g0) at (c15) {}; 
                                    \node[ellipse,draw=OliveGreen,dotted,minimum width=0.75cm,minimum height=0.75cm,line width=2pt] (g0) at (c21) {}; 
            
                                    \path[draw=blue,style=->] (0, -0.25) -- (2, -0.25) -- (4, -1.25) -- (6, -1.25) -- (8, -0.25) -- (10, -0.25);
                                \end{tikzpicture}
                            }
                            \caption{A path passing through two commits, one of which is a merge commit.}
                        \end{subfigure} 
            
                        \begin{subfigure}[t]{0.49\textwidth}
                            \centering
                            \scalebox{0.75}{
                                \begin{tikzpicture}
                                    \pic[draw=black] at (0, 0) {commitpath=1};
            
                                    \node[ellipse,draw=OliveGreen,dotted,minimum width=0.75cm,minimum height=0.75cm,line width=2pt] (g0) at (c13) {}; 
                                    \node[ellipse,draw=OliveGreen,dotted,minimum width=0.75cm,minimum height=0.75cm,line width=2pt] (g0) at (c22) {}; 
                                \end{tikzpicture}
                            }
                            \caption{Two commits in parallel branches, which violates the path requirement.}
                        \end{subfigure}\quad 
                        \begin{subfigure}[t]{0.49\textwidth}
                            \scalebox{0.75}{
                                \begin{tikzpicture}
                                    \pic[draw=black] at (0, 0) {commitpath=1};
            
                                    \node[ellipse,draw=OliveGreen,dotted,minimum width=0.75cm,minimum height=0.75cm,line width=2pt] (g0) at (c13) {}; 
                                    \node[ellipse,draw=OliveGreen,dotted,minimum width=0.75cm,minimum height=0.75cm,line width=2pt] (g0) at (c15) {}; 
                                \end{tikzpicture}
                            }
                            \caption{Two commits which violate the path requirement, because a path through the two commits would not pass through the branch being merged into.}
                        \end{subfigure} 
                        
                        \caption{Examples of configurations of linked commits that adhere to or violate the path requirement. Linked commits are circled in green.}
                        \label{fig:weak-branch-consistency}
                    \end{figure*}
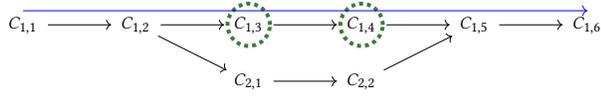
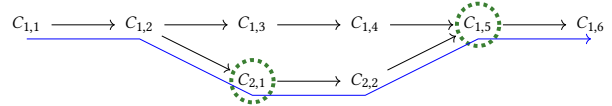
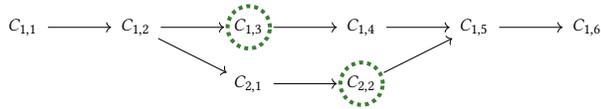
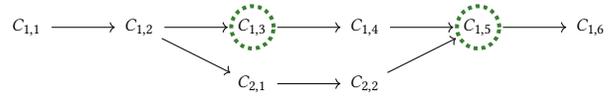

            \paragraph{Path Requirement} Some issues are mentioned in multiple commits. In this case, we assume that all these commits contain changes relevant to resolve the issue. However, in order to non-ambiguously handle these cases, we require that all commits mentioning a particular issue are contained in a single path from the root commit of the repository to the current HEAD commit; this means that at no point in time, the same issue is being worked on in parallel branches. One thing to note here is that whenever a merge commit is linked a specific issue, the path from root to HEAD must pass through the branch being merged into the other branch; The rationale here is that the branch must contain work relevant to the issue since developers tend to create branches for new work, which are later merged back. Figure~\ref{fig:weak-branch-consistency} illustrates the requirements more clearly with a number of examples. When the commits linked to an issue do not adhere to the path requirement, we ignore those commit and that particular issue.
    
                \paragraph{Merge Commit Disambiguation} Next, we remove possible ambiguity from merge commits. Specifically, we found that merge commits may introduce ambiguity or false positives if the issues linked to by the merge commit do not correspond perfectly with the issues linked to by the commits in the branch. For example, if a branch contains (separate) commits linked to two different issues, and the merge commit is linked to both, all changes for both issues will effectively become linked to both issues; This is because the merge commit effectively contains all the changes from all the commits in the branch.
    
                The naive fix would be to ignore merge commits in their entirety for the purposes of label extraction. However, we found this to be overly restrictive. In particular, it may occur that a branch contains only unlinked commits, which are then merged using a merge commit which does link to an issue. In this specific case, ignoring the merge commit unnecessarily ignores the particular issue. 
    
                Hence, in the end, we perform the following step to disambiguate merge commits:
    
                \begin{itemize}
                    \item If none of the commits in the branch is linked to an issue, the merge commit keeps it links.
                    \item If any of the commits in the branch is linked to an issue, the links of the merge commit are discarded. Any issues that were linked to by the merge commit, but not any commits in the branch, are discarded; there is no representative commit in the branch for this issue any more, and we no longer have a complete picture of the changes required to resolve such issues.
                \end{itemize}
    
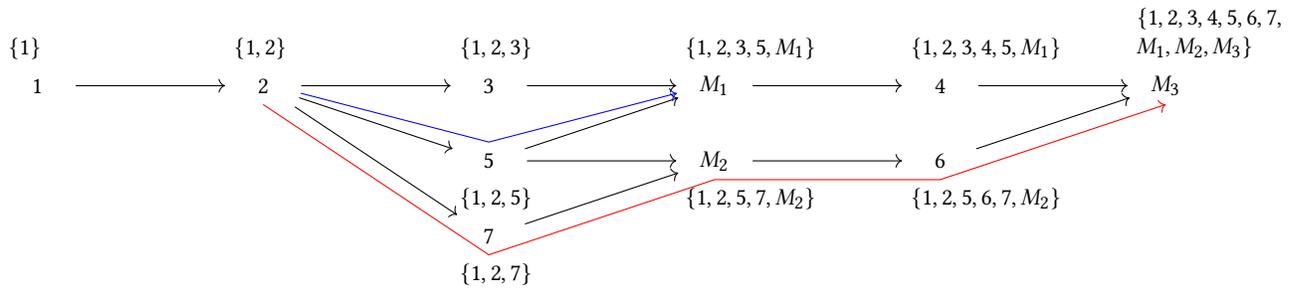
\begin{figure*}
                    \centering
                    \begin{tikzpicture}
                        \node[circle,minimum width=1cm] (c11) at (0, 0) {1};
                        \node[circle,minimum width=1cm] (c12) at (3, 0) {2};
                        \node[circle,minimum width=1cm] (c13) at (6, 0) {3};
                        \node[circle,minimum width=1cm] (c14) at (9, 0) {$M_1$};
                        \node[circle,minimum width=1cm] (c15) at (12, 0) {4};
                        \node[circle,minimum width=1cm] (c16) at (15, 0) {$M_3$};
    
                        \node[circle,minimum width=1cm] (c21) at (6, -1) {5};
                        \node[circle,minimum width=1cm] (c22) at (9, -1) {$M_2$};
                        \node[circle,minimum width=1cm] (c23) at (12, -1) {6};
    
                        \node[circle,minimum width=1cm] (c31) at (6, -2) {7};
    
                        \draw[->] (c11) to (c12);
                        \draw[->] (c12) to (c13);
                        \draw[->] (c13) to (c14);
                        \draw[->] (c14) to (c15);
                        \draw[->] (c15) to (c16);
    
                        \draw[->] (c12) to (c21);
                        \draw[->] (c21) to (c22);
                        \draw[->] (c22) to (c23);
    
                        \draw[->] (c12) to (c31);
    
                        \draw[->] (c21) to (c14);
                        \draw[->] (c23) to (c16);
                        \draw[->] (c31) to (c22);
    
                        \node[rectangle,anchor=west] (f11) at ([shift=({-0.5,0.5})]c11) {$\{1\}$};
                        \node[rectangle,anchor=west] (f12) at ([shift=({-0.5,0.5})]c12) {$\{1, 2\}$};
                        \node[rectangle,anchor=west] (f13) at ([shift=({-0.5,0.5})]c13) {$\{1, 2, 3\}$};
                        \node[rectangle,anchor=west] (f14) at ([shift=({-0.5,0.5})]c14) {$\{1, 2, 3, 5, M_1\}$};
                        \node[rectangle,anchor=west] (f15) at ([shift=({-0.5,0.5})]c15) {$\{1, 2, 3, 4, 5, M_1\}$};
                        \node[rectangle,anchor=west,align=left] (f16) at ([shift=({-0.5,0.7})]c16) {$\{1, 2, 3, 4, 5, 6, 7,$\\ $M_1, M_2, M_3\}$};
    
                        \node[rectangle,anchor=west] (f21) at ([shift=({-0.5,-0.5})]c21) {$\{1, 2, 5\}$};
                        \node[rectangle,anchor=west] (f22) at ([shift=({-0.5,-0.5})]c22) {$\{1, 2, 5, 7, M_2\}$};
                        \node[rectangle,anchor=west] (f23) at ([shift=({-0.5,-0.5})]c23) {$\{1, 2, 5, 6, 7, M_2\}$};
    
                        \node[rectangle,anchor=west] (f31) at ([shift=({-0.5,-0.5})]c31) {$\{1, 2, 7\}$};
    
    
    
                        \path[draw=red,style=->] (3, -0.25) -- (6, -2.25) -- (9, -1.25) -- (12, -1.25) -- (15, -0.25);
    
                        \path[draw=blue,style=->] (3.5, -0.1) -- (6, -0.75) -- (8.5, -0.1);
                    \end{tikzpicture}
                    \caption{Example of how the merge commit a priori file resolution algorithm works.}
                    \label{fig:single-entry-example}
                \end{figure*}

                \paragraph{Merge Commit A Priori Files} When predicting which files are going to change as a result of a given issue, we need the a priori state of the repository to make predictions on. For non-merge commits, this is just the state of the repository at their parent commit. However, this does not hold for merge commits; merge commits are a union of the commits in their corresponding branch. As such, the correct a priori situation would be the commit where the branch branches off from. However, because of arbitrary branching practices, such a commit may not always be unique (see Figure \ref{fig:single-entry-example}). As such, we make sure that for every linked merge commit, there is only a single unique ``entry point'' into the branch. If this is not the case, the issues associated with the merge commit are discarded from the dataset.
    
                How we determine this, is illustrated in Figure~\ref{fig:single-entry-example}. Starting from the root commit(s), for every commit, we compute the set of commits that (topologically) came before it. When a branch with commits $B$ is merged into a branch with commits $A$, we consider the branch with the collection of commits $B \setminus A$ to be the actual branch being merged. For simplicity, we assume 2-way merge commits only here, and do not support arbitrary $N$-way merge commits\footnote{In practice, we only encountered 2-way merge commits.}. For every merge commit with linked issues remaining after the merge disambiguation step, we check the parents of all the issues in its corresponding branch; if there is exactly one commit with exactly one parent not contained in the branch, then the branch satisfies our unique entry point requirement. In Figure~\ref{fig:single-entry-example}, merge commit $M_1$ satisfies the requirement, while $M_3$ does not; commit $M_2$ has parent $5$ which is not contained in $\{1, 2, 5, 6, 7, M_2\} \setminus \{1, 2, 3, 4, 5, M1\} = \{6, 7, M_2\}$.
    
        \subsubsection{Labelled Sample Generation}
            In this section, we describe how we use the obtained links to generate labelled datasets.             
            To extract labels, we consider individual commits. Given an issue $I$ and commit $C$, we perform the following steps to come up with a set of labelled samples:

            \begin{enumerate}
                \item Obtain the set of modified files by diffing the commit against its parent. For merge commits, we diff against the first parent -- which represents the branch being merged into. Due to the fact that creating a diff is ambiguous for a general $N$-way merge with $N > 2$ commits, we restrict ourselves to 2-way merge commits only. 

                \item Get the a priori state $S$ of the project before $C$ is merged. More details are given in Section~\ref{sec:datast-2}.

                \item We filter the files in the repository based on extension so that only source code files are kept (e.g. we exclude documentation, configuration, and README files). Every source code file $s_i \in S$ which is modified or removed in $C$ is considered a positive sample for the pair $(I, C)$\footnote{In some circumstances, a file may be deleted and re-added in the same commit. These files are included as positive samples.}; All other files $s_i \in S$ are considered as negative samples. In the rare occasion where an existing non-code file is renamed to a file name with source code extension, we count it as a new file being added and the old file is not included as a positive nor a negative sample. In case no positive samples exists (additions only or all modified files are not source code files), we move on to the next commit linked to the issue. If no commit linked to the issue has a non-empty set of positive samples, the issue is discarded from the dataset. 
            \end{enumerate}

            In Table~\ref{tab:dataset-overview}, we provide an overview of the dataset after applying these steps.

            \begin{table*}[]
                \centering
                \caption{Overview of our dataset (for the \textit{First Commit Only} variant).}
                \label{tab:dataset-overview}
                \begin{tabular}{
                    p{1.5cm} 
                    |
                    >{\centering\arraybackslash}p{1.2cm} 
                    >{\centering\arraybackslash}p{1.8cm}
                    >{\centering\arraybackslash}p{1cm} 
                    >{\centering\arraybackslash}p{1cm}
                    |
                    >{\centering\arraybackslash}p{0.8cm} 
                    >{\centering\arraybackslash}p{0.8cm}
                    >{\centering\arraybackslash}p{0.8cm} 
                    >{\centering\arraybackslash}p{0.9cm}
                    |
                    >{\centering\arraybackslash}p{0.8cm} 
                    >{\centering\arraybackslash}p{0.8cm}
                    >{\centering\arraybackslash}p{0.8cm} 
                    >{\centering\arraybackslash}p{0.9cm}
                }
                     \toprule
                    \multirow{2}{1.5cm}{Project} & \multirow{2}{1.2cm}{\centering Unknown Issues} & \multirow{2}{1.8cm}{\centering Max Linkable} & \multirow{2}{1cm}{\centering Linked} & \multirow{2}{1cm}{\centering In Dataset} & \multicolumn{4}{c|}{\# Files/Issue}  & \multicolumn{4}{c}{\# Positives/Issue} \\ 
                    & & & & & Min      & Max      & Mean         & Median       & Min      & Max      & Mean     & Median      \\ \midrule
                    Avro                     & 4                    & 2452 (59.3\%)     & 2447         & 1936             & 83       & 1416     & 894.1        & 994          & 1        & 1048     & 5.6      & 2           \\
                    Maven                    & 191                  & 2798 (37.9\%)     & 2199         & 1637             & 240      & 2954     & 1065.0       & 892          & 1        & 1420     & 8.1      & 2           \\
                    Thrift                   & 3                    & 3645 (62.1\%)     & 3591         & 2630             & 334      & 1402     & 928.8        & 942.5        & 1        & 361      & 3.8      & 1           \\
                    Tika                     & 1                    & 2557 (57.7\%)     & 2530         & 2008             & 46       & 1817     & 895.7        & 845          & 1        & 623      & 5.6      & 2           \\
                    TomEE                    & 11                   & 2181 (54.0\%)     & 2123         & 1428             & 3425     & 6626     & 5221.8       & 5216         & 1        & 1572     & 4.3      & 2           \\
                    Spring Roo               & 0                    & 2304 (57.8\%)     & 2263         & 1727             & 311      & 1145     & 719.6        & 708          & 1        & 1015     & 4.9      & 1           \\
                    Spring Data MongoDB      & 2                    & 1337 (51.6\%)     & 1337         & 972              & 283      & 958      & 601.7        & 563          & 1        & 853      & 7.2      & 2           \\
                    \bottomrule

                \end{tabular}

                Unknown issues are issues in commits which were not present in the issue tracker; Max Linkable is the number of issues linked to from commits; Linked is the number of issues linked to after applying the refinement steps from Section \ref{sec:datast-2}; In Dataset represents the number of linked issues after label extraction.
            \end{table*}

    \subsection{Features}
        The pre-processing is based on \cite{maarleveld_maestro_2024}. Specifically, for issues, we first removed any Jira-specific formatting syntax\footnote{\url{https://jira.atlassian.com/secure/WikiRendererHelpAction.jspa?section=all}}. For text modifiers (e.g. headings, text effects, colours, tables, and lists), we only removed the formatting syntax but kept the actual text. Images, attachments, and hyperlinks were removed in their entirety. We also removed non-text formatting (e.g. horizontal rules, graphical emoticons). For code blocks, ``noformat'' blocks, and panels, we experimented with multiple options: 1) removing only the formatting but keeping the content, 2) removing them including content, and 3) removing them, including content, but inserting a special marker word.

        Furthermore, we also experimented with traditional natural language processing feature preprocessing techniques. We applied these to both the issues and the source files. Specifically, we experimented with lower-casing and stemming. We also experimented with splitting lowerCamelCase and UpperCamelCase words into their constituent sub-tokens \cite{dit_can_2011}.
        
    \subsection{Retrieval Models}
        For this study, we compare a number of text retrieval models, which we explain in more detail below. In the following section, $f(w, D)$ denotes the number of occurrences of the term $w$ in document $D$, $n(w)$ denotes the number of documents containing the term $w$, and $N$ denotes the total number of documents.

        \subsubsection{Vector Space Model with TF-IDF}
            TF-IDF (Term Frequency - Inverse Document Frequency) is a way of assigning scores to terms in documents to indicate their importance where, for each term in a document, its frequency is multiplied with the inverse document frequency of that term \cite{manning_introduction_to_information_retrieval_2008}.  Note that we follow scikit-learns implementation of smooth IDF (Eq. \ref{eq:tfidf-1}), where we act as if a single document containing all terms is added to the corpus in order to avoid zero divisions\footnote{\url{https://scikit-learn.org/stable/modules/feature\_extraction.html\#tfidf-term-weighting}}. 
            


            \begin{equation}\label{eq:tfidf-1}
                \text{IDF}_1(w) = \log\left(\frac{N + 1}{n(w) + 1}\right) + 1
            \end{equation}

            In the vector space model (VSM), documents are encoded as vectors of TF-IDF scores, where each dimension represents a specific word. For information retrieval, the query (in our case, an issue), is also encoded as a vector, and the documents are ranked according to their similarity with the query, where similarity is measured using cosine similarity \cite{manning_introduction_to_information_retrieval_2008}.

        \subsubsection{Latent Semantic Indexing}
            Latent Semantic Indexing (LSI) is the process of using Singular Value Decomposition (SVD) to convert vectors used in the vector space model into a lower dimensional representation before computing the cosine similarity. This has been shown to sometimes result in improved precision \cite{manning_introduction_to_information_retrieval_2008}.

        \subsubsection{Revised Vector Space Model}
            The Revised Vector Space Model (rVSM) is a variant of the vector space model which was adapted for improved performance in bug localisation \cite{zhou_where_2012}. It comes with two major improvements over the traditional vector space model. First of all, it uses $\log(f(w, D) + 1)$ instead of $f(w, D)/|D|$ for document frequency. This prevents documents from becoming artificially more important because of inflated usage of a specific term. Second, for a given query $Q$ and document $D$, with vectors $V_Q$ and $V_D$, the similarity computation is changed to the one show in Eq. \ref{eq:rvsm-1} ($x_{min}$ and $x_{max}$ represent minimum and maximum document lengths, respectively). The idea is that for bug localisation in particular, longer source files are more likely to contain bugs and should thus receive higher scores \cite{zhou_where_2012}. 

            \begin{equation}\label{eq:rvsm-1}
                \text{sim}(Q, D) = \frac{1}{1 + e^{-g(|D|)}}\frac{V_Q \cdot V_D}{|V_Q| |V_D|}
            \end{equation}

            \begin{equation}
                g(x) = \frac{x - x_{min}}{x_{max} - x_{min}}
            \end{equation}

        \subsubsection{BM25}
            BM25 (\cite{robertson1995okapi}; Eq. \ref{eq:bm25-1} -- Note that $\ell_{avg}$ denotes average documennt length) is a ranking function for computing a relevance score between a query and a document. It is conceptually similar to TF-IDF with log smoothing for term frequency, but uses different calculations; the calculation ensures diminishing increases in score as the number of occurrences of a term becomes larger. We also use an extension called BM25+ which involves adding the parameter $\delta$ term in Eq. \ref{eq:bm25-2} \cite{lv_lower-bounding_2011}. Furthermore, to enable matching on both file name and file content, we use an extension called BM25F which generalises BM25 to documents with multiple fields \cite{robertson_simple_2004}. It involves replacing $f(q, D)$ with the weighted sum of the term frequencies in all fields.
            
            \begin{equation}\label{eq:bm25-1}
                \text{BM25}(Q, D) = \sum_{q_i \in Q} \text{IDF}_2(q_i) \cdot s(q_i, D)
            \end{equation}

            \begin{equation}\label{eq:bm25-2}
                s(q, D) = \frac{(k_1 + 1)f(q, D))}{f(q, D) + k_1\left(1 - b + b\frac{|D|}{\ell_{avg}}\right)} + \delta
            \end{equation}

            \begin{equation}\label{eq:bm25-3}
                \text{IDF}_2(q_i) = \ln\left(\frac{N - n(q_i) + 0.5}{n(q_i) + 0.5} + 1\right)
            \end{equation}

    \subsection{Evaluation Metrics}
        We argue that for this particular problem, accurate highly ranked results are more important than capturing all the results. The reason for this is that with an initial seed set, recall-focussed change impact analysis techniques can be applied to further complete the set. The primary metrics we used are $\text{hit}@k$ (Eq \ref{eq:hit-at-k}), $\text{precision}@k$ (Eq. \ref{eq:precision-at-k}), Mean Reciprocal Rank (MRR) (Eq. \ref{eq:mrr}), and R-Precision (which is similar to $\text{precision}@k$, but for every query, $k$ is set to the number of positive samples for that specific query). For completeness, we also computed $\text{recall}@k$ (Eq. \ref{eq:recall-at-k}). In Equations \ref{eq:hit-at-k}-\ref{eq:mrr}, $\text{tp}_i@k$, $\text{fp}_i@k$, and $\text{fn}_i@k$ denote the true positive, false positive, and false negative counts within the top $k$ ranked samples for issue $i$. $\text{rank}_i$ denotes the rank of the first positive sample. $I$ denotes the complete set of issues, and $[\cdot]$ denotes Iverson brackets. We computed each metric for all $k \in \{1, 5, 10\}$. We abbreviate $\text{precision}@k$ as $P@k$, $\text{hit}@k$ as $H@k$, $\text{recall}@k$ as $R@k$, and R-Precision as RP.

        \begin{equation}\label{eq:hit-at-k}
            \text{hit}@k = \frac{1}{|I|}\sum_{i \in I} [\text{tp}_i@k \ge 1]
        \end{equation}

        \begin{equation}\label{eq:precision-at-k}
            \text{precision}@k = \frac{1}{|I|}\sum_{i \in I}\frac{\text{tp}_i@k}{\text{tp}_i@k + \text{fp}_i@k}
        \end{equation}

        \begin{equation}\label{eq:recall-at-k}
            \text{recall}@k = \frac{1}{|I|}\sum_{i \in I}\frac{\text{tp}_i@k}{\text{tp}_i@k + \text{fn}_i@k}
        \end{equation}

        \begin{equation}\label{eq:mrr}
            \text{MRR} = \frac{1}{|I|}\sum_{i \in I} \frac{1}{\text{rank}_i} 
        \end{equation}

    \subsection{Experiments}
    
        \subsubsection{RQ 1: Performance}\label{sec:exp-rq1}
            To answer RQ 1, we performed a number of experiments with a number of different retrieval models. Specifically, the models we used are VSM, LSI with 500 dimensional vectors (LSI-500), LSI with 1000 dimensional vectors (LSI-1000), rVSM, and BM25.

            We split our data into a validation set and a test set using a 50/50 split (with respect to the number of issues). Following best practices for dealing with the temporal data typically encountered in traceability problems (\cite{gama_towards_2024,hirsch_best_2025,izadi_empirical_2023}), the split was done in a temporal fashion; issues were ordered according to their first commit, and then a 50/50 split was made\footnote{Technically, such a split is less of a concern in this scenario because we have little to no trainable components in our method.}. The validation set is used for model selection (preprocessing settings), while the test set is used for all other performance evaluations. 

            For every experiment or evaluation we perform, we use the same procedure. We consider each issue in the split one by one. For each issue, we get its associated source code files, and use these documents (both their file name and content) to compute IDF weights. For VSM, rVSM, and both LSI models, we concatenate the file name with the file content. For BM25, we treat the two as two fields of a single document.
            
            We experimented with a number of different configurations for the preprocessing. Specifically, we first evaluated all five models with different formatting removal techniques (raw text, formatting removal, block removal, and replacing blocks with markers). For these experiments, we fixed the other preprocessing (convert all text to lower casing, no stemming, and no sub-token splitting). Furthermore, for BM25, we chose the parameters $k_1 = 1.2$, $b = 0.75$, $\delta = 1.0$ (following recommended practices \cite{manning_introduction_to_information_retrieval_2008,lv_lower-bounding_2011}), and equal weights of $1$ for the file name and file content. We found that not removing any formatting resulted in the best performance (results available in our replication package \cite{study-replication-package}). After this, we optimised the remaining preprocessing options, where we experimented with i) no lower casing, no stemming, ii) just lower casing, and iii) lower casing and stemming. We tried all of these with and without sub-token splitting, for a total of six different combinations. Lower casing and stemming resulted in the best performance. Omitting sub-token splitting resulted in better top-1 performance for all metrics, while enabling sub-token splitting results in better top-5 and top-10 performance for all metrics (results available in our replication package \cite{study-replication-package}). Since our main focus is correct highly ranked files, we focus our analysis on the variant without sub-token splitting. In our replication package, we also included the results for the variant with sub-token splitting.

        \subsubsection{RQ 2: Issue Type}
            To investigate the effect of issue type, we performed two types of tests: 1) experiments where our dataset only contains issues of a single type, and 2) experiments where we leave issues of a single type out. We will present the results for the best performing model from RQ 1, but results for all other methods are included in the replication package \cite{study-replication-package}. We use the Kruskal-Wallis test with threshold $\alpha = 0.05$ to determine whether a statistically significant difference between issue types exists. For cases where it does, we use Conover's post-hoc test to perform pairwise comparisons. We exclude the hit@$k$ metrics from this analysis, since these are binary (not ratio) variables. 

            We determine issue type based on the ``issue type'' field from Jira. Because some projects used slightly different types, we consolidated issue types into a number of overarching categories. This categorisation was done by the first author, and double-checked by the 3rd author. Issue types or categories with less than five issues on average over all projects were not considered for exclusion. In the end, we arrived at the categories \textit{Bug}, \textit{New Feature}, \textit{Improvement}, and \textit{Task}.

        \subsubsection{RQ 3: Identifiers and File Names}
            To investigate the effect of identifiers and file names, we take the predictions of our retrieval models for each individual issue. For each issue, we use regex to extract and count the number of identifiers and file names. Next, for each performance metric we record, we compute the Spearman correlation between the total number of identifiers and file names and the performance metric. Moreover, we also do this separately per issue type in order to investigate whether there are differences between issue types. Again, we present the results for the best-performing model from RQ1, but results for all other models are once again included in our replication package \cite{study-replication-package}. We once again exclude the hit@$k$ metrics from our analysis. 

\section{Results}\label{sec:results}
    \subsection{RQ 1: Performance}
        In Table \ref{tab:performance-per-method}, the average performance of every information retrieval method over all projects in our dataset is presented, with the best score per metric \underline{underlined}. BM25 is the best performing model according to every metric. We also observe that generally, TF-IDF outperforms LSI-1000, which in turn outperforms LSI-500. This suggests that for this particular problem and dataset, reducing the dimensionality does not improve performance; performance becomes worse as more information is discarded. Finally, we note that rVSM is the worst performing model. This shows that in this case, bug-specific modifications to TF-IDF may actually lead to worse performance when considering non-bug issues.

        Table \ref{tab:performance-per-project} shows the average performance for all projects in the dataset (with best scores once again \underline{underlined}). The table shows that there are considerable performance differences between the different datasets. Hence, results on one project do not provide performance guarantees for other projects. 

        \begin{tcolorbox}[boxsep=1pt,left=10pt,right=10pt,top=3pt,bottom=3pt]
            \textbf{RQ 1 key takeaways}: \\ 
            $\bullet$ BM25 is the best performing method. \\
            $\bullet$ The bug-specific rVSM method performs worst. \\
            $\bullet$ Performance is strongly project-dependent.
        \end{tcolorbox}
        
        \begin{table}[]
            \centering
            \caption{Average performance of all retrieval methods over all projects in the dataset.}
            \label{tab:performance-per-method}
            \begin{tabular}{
                p{1cm} 
                >{\centering\arraybackslash}p{0.74cm}
                >{\centering\arraybackslash}p{0.74cm}
                >{\centering\arraybackslash}p{0.74cm}
                >{\centering\arraybackslash}p{0.74cm}
                >{\centering\arraybackslash}p{0.74cm}
                >{\centering\arraybackslash}p{0.74cm}
                >{\centering\arraybackslash}p{0.74cm}
            }
                \toprule
                Metric           & TF-IDF       & rVSM     & LSI-500      & LSI-1000     & BM25                 & Mean        \\ \midrule
                P@1              & 0.302        & 0.243    & 0.273        & 0.295        & \underline{0.321}    & 0.287       \\
                P@5              & 0.150        & 0.136    & 0.142        & 0.149        & \underline{0.164}    & 0.148       \\
                P@10             & 0.105        & 0.097    & 0.102        & 0.105        & \underline{0.114}    & 0.105       \\
                H@5              & 0.517        & 0.477    & 0.484        & 0.515        & \underline{0.573}    & 0.513       \\
                H@10             & 0.604        & 0.575    & 0.578        & 0.605        & \underline{0.666}    & 0.606       \\
                R@1              & 0.175        & 0.140    & 0.153        & 0.169        & \underline{0.184}    & 0.164       \\
                R@5              & 0.338        & 0.312    & 0.314        & 0.336        & \underline{0.381}    & 0.336       \\
                R@10             & 0.422        & 0.398    & 0.402        & 0.420        & \underline{0.471}    & 0.423       \\
                RP               & 0.253        & 0.213    & 0.230        & 0.247        & \underline{0.265}    & 0.242       \\
                MRR              & 0.405        & 0.356    & 0.376        & 0.400        & \underline{0.438}    & 0.395       \\
                \bottomrule
            \end{tabular}

        \end{table}

        \begin{table}[]
            \centering
            \caption{Average performance for each dataset, over all retrieval methods.}
            \label{tab:performance-per-project}
            \begin{tabular}{
                p{0.8cm} 
                >{\centering\arraybackslash}p{0.69cm}
                >{\centering\arraybackslash}p{0.69cm}
                >{\centering\arraybackslash}p{0.69cm}
                >{\centering\arraybackslash}p{0.69cm}
                >{\centering\arraybackslash}p{0.69cm}
                >{\centering\arraybackslash}p{0.69cm}
                >{\centering\arraybackslash}p{0.69cm}
            }
               \toprule
                Metric           & Avro                 & Maven    & Tika                 & Thrift               & TomEE    & DMDB      & Roo      \\ \midrule
                P@1      & 0.340                & 0.194    & \underline{0.360}    & 0.360                & 0.190    & 0.316                    & 0.244           \\
                P@5      & 0.173                & 0.109    & \underline{0.193}    & 0.151                & 0.096    & 0.168                    & 0.147           \\
                P@10     & 0.118                & 0.081    & \underline{0.136}    & 0.099                & 0.065    & 0.124                    & 0.109           \\
                H@5            & \underline{0.608}    & 0.357    & 0.608                & 0.589                & 0.373    & 0.567                    & 0.489           \\
                H@10           & \underline{0.696}    & 0.422    & 0.692                & 0.691                & 0.455    & 0.684                    & 0.600           \\
                R@1         & 0.202                & 0.106    & 0.174                & \underline{0.267}    & 0.123    & 0.147                    & 0.130           \\
                R@5         & 0.409                & 0.212    & 0.373                & \underline{0.463}    & 0.267    & 0.311                    & 0.318           \\
                R@10        & 0.503                & 0.266    & 0.460                & \underline{0.553}    & 0.338    & 0.414                    & 0.426           \\
                RP      & 0.293                & 0.168    & 0.284                & \underline{0.325}    & 0.165    & 0.252                    & 0.205           \\
                MRR              & 0.462                & 0.275    & \underline{0.476}    & 0.471                & 0.281    & 0.438                    & 0.360           \\
                \bottomrule
            \end{tabular}

        \end{table}

    \subsection{RQ 2: Issue Type}
\begin{figure}
            \centering
            \includegraphics[width=\linewidth]{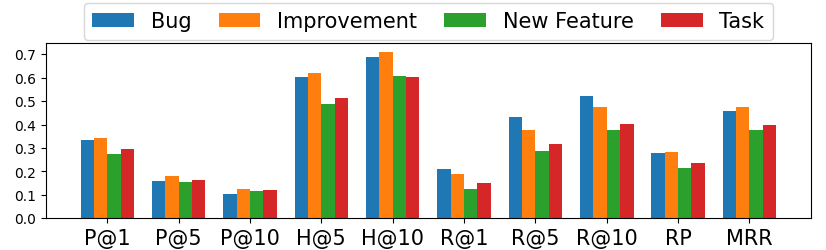}
            \caption{Average performance per issue type of BM25 over all projects in the dataset.}
            \label{fig:bm25-issue-types-specific}
        \end{figure}

\begin{figure}
            \centering
            \includegraphics[width=\linewidth]{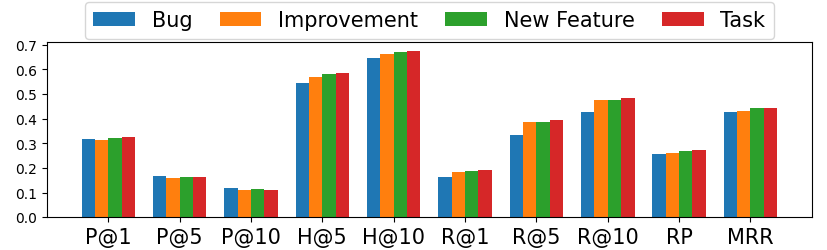}
            \caption{Average performance per held-out issue type of BM25 over all projects in the dataset.}
            \label{fig:bm25-issue-types-holdout}
        \end{figure}

        In Tables \ref{tab:issue-type-kw-specific} we present the results of comparing the performance scores of BM25 on the four major categories of issue types using the Kruskal-Wallis test. Table \ref{tab:issue-type-kw-holdout} shows similar results, but for the case where we held one issue type out of the dataset and evaluated on all others. In both tables, we recorded effect size ($\epsilon^2$), and marked statistically significant results with an asterisks (*). \underline{Underlined} results represent the highest effect sizes (which are small -- $\epsilon^2 \in [0.01, 0.06)$), while normal text represents negligible effect size ($\epsilon^2 < 0.01$). Figures \ref{fig:bm25-issue-types-specific} and \ref{fig:bm25-issue-types-holdout} also show the performance scores of BM25 per included and held out issue type. From Figure~\ref{fig:bm25-issue-types-specific} and Table~\ref{tab:issue-type-kw-specific}, we observe that there are variations between issue types, and that these variations are statistically significant for a number of combinations of project and metric. However, these effect seems to be project-dependent, and affect recall more widely than other metrics. Using Conover's post-hoc test (results in replication package \cite{study-replication-package}), we find that generally, statistically significant results vary in direction for \textit{bug} and \textit{improvement} issue types. \textit{Improvement} and \textit{bug} are statistically higher than \textit{new feature}. \textit{Improvement} is also statistically higher than \textit{task}. \textit{Bug \& Task}, and \textit{Task \& New Feature}, have no statistically significant pairwise differences. Although these tendencies exists, results are project-specific, and for some directions the inter-type differences are flipped. 
        
        From Figure \ref{fig:bm25-issue-types-holdout} and Table \ref{tab:issue-type-kw-holdout}, we observe that removing an issue type has less impact; there are less significant performance differences, and effect sizes are smaller. Hence, we conclude that there are significant differences in performances between issue types, but there is not a single issue type that is detrimental to good performance. The results of Conover's post-hoc test are consistent with our findings from Conover's test for the isolated issue types.

        \begin{tcolorbox}[boxsep=1pt,left=10pt,right=10pt,top=3pt,bottom=3pt]
            \textbf{RQ 2 key takeaways}: \\ 
            $\bullet$ The effect of issue type is frequently statistically significant, with small to negligible effect size. \\
            $\bullet$ No particular issue type is detrimental to performance. \\
            $\bullet$ The exact effects of issue type are project-dependent.
        \end{tcolorbox}
        
        \begin{table}[]
            \centering
            \caption{Result of the Kruskal-Wallis Test (effect size) on BM25 performance per issue type.}
            \label{tab:issue-type-kw-specific}
            \begin{tabular}{
                p{0.8cm} 
                >{\centering\arraybackslash}p{0.69cm}
                >{\centering\arraybackslash}p{0.69cm}
                >{\centering\arraybackslash}p{0.69cm}
                >{\centering\arraybackslash}p{0.69cm}
                >{\centering\arraybackslash}p{0.69cm}
                >{\centering\arraybackslash}p{0.69cm}
                >{\centering\arraybackslash}p{0.69cm}
            }
                \toprule
                Metric           & Avro                     & Maven                    & Tika                     & Thrift                   & TomEE                    & DMDB      & Roo      \\ \midrule
                P@1      & 0.004                    & 0.006*                   & 0.005*                   & 0.006*                   & 0.000                    & 0.008                    & 0.000           \\
                P@5      & 0.004                    & \underline{0.015}*       & 0.001                    & \underline{0.010}*       & 0.008*                   & 0.001                    & 0.009*          \\
                P@10     & 0.006*                   & \underline{0.019}*       & 0.002                    & 0.000                    & 0.005                    & 0.003                    & 0.005           \\
                R@1         & 0.008*                   & 0.006                    & \underline{0.019}*       & \underline{0.012}*       & 0.000                    & 0.010                    & 0.000           \\
                R@5         & \underline{0.019}*       & \underline{0.012}*       & \underline{0.038}*       & \underline{0.048}*       & \underline{0.029}*       & 0.000                    & 0.002           \\
                R@10        & \underline{0.024}*       & \underline{0.011}*       & \underline{0.034}*       & \underline{0.037}*       & \underline{0.031}*       & 0.000                    & 0.000           \\
                RP      & 0.010*                   & 0.010*                   & 0.004                    & 0.007*                   & 0.002                    & 0.007                    & 0.000           \\
                MRR              & \underline{0.013}*       & \underline{0.027}*       & 0.010*                   & \underline{0.012}*       & \underline{0.018}*       & 0.004                    & 0.003           \\
                \bottomrule
            \end{tabular}
        \end{table}

        \begin{table}[]
            \centering
            \caption{Result of the Kruskal-Wallis Test (effect size) on BM25 performance per held-out issue type.}
            \label{tab:issue-type-kw-holdout}
            \begin{tabular}{
                p{0.8cm} 
                >{\centering\arraybackslash}p{0.69cm}
                >{\centering\arraybackslash}p{0.69cm}
                >{\centering\arraybackslash}p{0.69cm}
                >{\centering\arraybackslash}p{0.69cm}
                >{\centering\arraybackslash}p{0.69cm}
                >{\centering\arraybackslash}p{0.69cm}
                >{\centering\arraybackslash}p{0.69cm}
            }
               \toprule
                Metric           & Avro     & Maven        & Tika         & Thrift                   & Tomee        & DMDB      & Roo      \\ \midrule
                P@1      & 0.000    & 0.000        & 0.000        & 0.001                    & 0.000        & 0.001                    & 0.000           \\
                P@5      & 0.000    & 0.001        & 0.000        & 0.002*                   & 0.000        & 0.000                    & 0.000           \\
                P@10     & 0.000    & 0.002*       & 0.000        & 0.000                    & 0.000        & 0.000                    & 0.000           \\
                R@1         & 0.000    & 0.000        & 0.003*       & 0.003*                   & 0.000        & 0.001                    & 0.000           \\
                R@5         & 0.001    & 0.000        & 0.006*       & \underline{0.012}*       & 0.003*       & 0.000                    & 0.000           \\
                R@10        & 0.001    & 0.001        & 0.005*       & 0.009*                   & 0.003*       & 0.000                    & 0.000           \\
                RP      & 0.000    & 0.001        & 0.000        & 0.002*                   & 0.000        & 0.000                    & 0.000           \\
                MRR              & 0.000    & 0.004*       & 0.001        & 0.003*                   & 0.001        & 0.000                    & 0.000           \\
                \bottomrule
            \end{tabular}
        \end{table}

    \subsection{RQ 3: Number of Identifiers and File Names}
\begin{figure*}
            \centering
            \includegraphics[width=\textwidth]{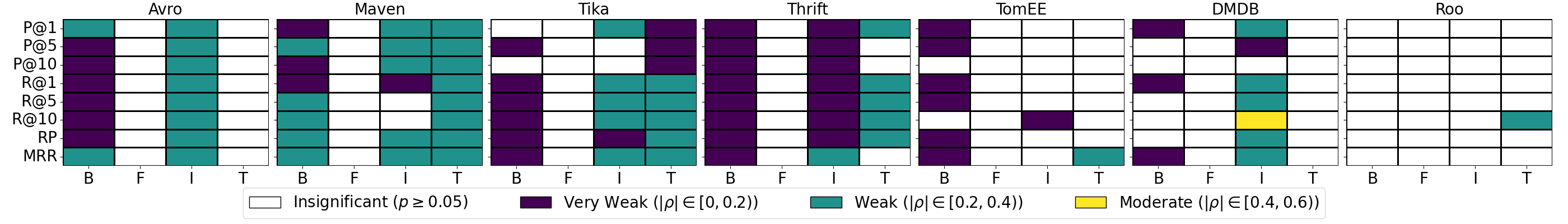}
            \caption{Spearman correlation per issue type (B -- Bugs / F -- New Feature / I -- Improvement / Task -- Task) between the different performance metrics and the number of identifiers and file names.}
            \label{fig:correlation-by-type}
        \end{figure*}

        In Table~\ref{tab:identifier-corrlation-joint}, we present the Spearman correlations between the performance of BM25 and the number of identifiers and file names in issues. Results with asterisks (*) are statistically significant. \underline{Underlined} text denotes the highest correlations (which are weak -- $|\rho| \in [0.2, 0.4)$)); normal text denotes very weak/negligible correlation ($|\rho| \in [0, 0.2)$). The correlation is almost always significant, except for Spring Roo -- Once again showing that results may be strongly project-dependent. Consistent with \citet{widyasari_influence_2022}, we find that the effect size is generally small. Effect size seems to once again be somewhat project-dependent. In Figure~\ref{fig:correlation-by-type}, we present the Spearman correlation per issue type. From this, we observe that \textit{bugs} and \textit{improvements} issue types have the most consistent, statically significant correlations, followed by \textit{tasks}. The effect size seems to be greater for \textit{improvements} and \textit{tasks} than it is for \textit{bug}; this could suggest that the number of identifiers is not the only factor explaining differences between issue types. However, definitive explanations are difficult for \textit{task}, since its exact usage may be project-dependent. \textit{New features} never have a significant correlation, indicating that the presence of identifiers does not provide any insights. 

        \begin{tcolorbox}[boxsep=1pt,left=10pt,right=10pt,top=3pt,bottom=3pt]
            \textbf{RQ 3 key takeaways}: \\ 
            $\bullet$ Identifiers almost always positively influence performance. \\
            $\bullet$ Influence of identifiers varies per issue type. \\
            $\bullet$ The exact effect of identifiers is strongly project-dependent.
        \end{tcolorbox}

        \begin{table}[]
            \centering
            \caption{Spearman correlation between the number of identifiers \& file names and performance of BM25.}
            \label{tab:identifier-corrlation-joint}
            \begin{tabular}{
                p{0.8cm} 
                >{\centering\arraybackslash}p{0.69cm}
                >{\centering\arraybackslash}p{0.69cm}
                >{\centering\arraybackslash}p{0.69cm}
                >{\centering\arraybackslash}p{0.69cm}
                >{\centering\arraybackslash}p{0.69cm}
                >{\centering\arraybackslash}p{0.69cm}
                >{\centering\arraybackslash}p{0.69cm}
            }
                \toprule
                Metric           & Avro                 & Maven                & Tika                 & Thrift               & TomEE    & DMDB      & Roo      \\ \midrule
                P@1      & \underline{0.25}*    & 0.17*                & 0.18*                & 0.16*                & 0.13*    & 0.15*                    & -0.01           \\
                P@5      & \underline{0.22}*    & \underline{0.21}*    & 0.11*                & 0.18*                & 0.12*    & 0.07                     & -0.04           \\
                P@10     & 0.18*                & 0.18*                & 0.07*                & 0.15*                & 0.11*    & 0.07                     & -0.04           \\
                R@1         & \underline{0.25}*    & 0.17*                & \underline{0.23}*    & 0.16*                & 0.13*    & 0.17*                    & 0.01            \\
                R@5         & \underline{0.26}*    & \underline{0.21}*    & \underline{0.27}*    & \underline{0.20}*    & 0.12*    & 0.12*                    & 0.04            \\
                R@10        & \underline{0.26}*    & \underline{0.20}*    & \underline{0.26}*    & 0.19*                & 0.12*    & 0.17*                    & 0.06            \\
                RP      & \underline{0.24}*    & 0.17*                & 0.18*                & 0.15*                & 0.10*    & 0.11*                    & -0.02           \\
                MRR              & \underline{0.28}*    & \underline{0.22}*    & \underline{0.23}*    & \underline{0.20}*    & 0.15*    & 0.16*                    & -0.01           \\
                \bottomrule
            \end{tabular}
        \end{table}

\section{Discussion}\label{sec:discussion}

\subsection{Implications for Researchers \& Practitioners}
    In this work, we advocate for increased research attention for file localisation for issues, regardless of their type. Our work mainly has implications for researchers or practitioners working on file localisation models. We provide a dataset and reusable tooling for expanding or creating a new dataset. We also provide a set of baseline results to compare more complex models against. 
    
    Our results highlight the importance of evaluating methods on a wide variety of different projects, since performance can differ significantly from project to project; Factors affecting performance in one project may have little to no effect in others, or vice versa. 
    
    On the one hand, the poor performance of rVSM also highlights the need for the development for more general methods; bug-based heuristics may not generalise well to general issue types. On the other hand, our results show that general purpose methods may not be affected to a great degree by variations in issue type; We found no motivation to focus on or exclude certain types of issues. However, researchers \& practitioners should be wary that certain factors (\# identifiers in our case) might have different effects across issue types of projects. This motivates the development of new methods which can be adapted to individual projects or groups of projects. Examples include the development of machine learning models which can be fine-tuned on a project's own historical data in order to capture project-specific characteristics.
    
    All this also has implications for end users; The success of file localisation tools might heavily depend on project characteristics or development practices, and might depend on good and consistent practices with regards to the creation of new issues.

\subsection{Threats To Validity}\label{sec:threats-to-validity}
    \textbf{Construct Validity}: We assumed that, when a commit message mentions multiple issues, the modified files are valid for all issues. However, in practice, developers often include multiple unrelated or poorly related changes in a single commit \cite{Herzig2016}. In bug localisation research, this was found to not have a significant impact \cite{kochhar_potential_2014,widyasari_influence_2022}. Nevertheless, untangling of commits into independent change sets has seen active research interest (e.g. \cite{10.1145/3368089.3409693,10.1145/3691620.3694996,10.1145/3540250.3549171}). Hence, future work could look into combining commit untangling with our work presented here to investigate 1) the impact of false positives created by tangled commit, and 2) the performance implications from untangling commits before linking.
        

    We also assumed that all commits mentioning a single issue contain changes required to change that issue. However, that assumption does not always hold. Examples include revert commit (e.g. ``Revert KEY-12''), or issues building on other issues (e.g. ``KEY-23: Implement new feature using framework from KEY-20''). Additional heuristic or machine learning methods are required to detect and resolve such cases. 

    The requirement that all commits linked to a particular issue should be contained in a single path from root to HEAD, is more restrictive than it needs to be. In particular, cherry-picked commits can be safely ignored. We considered reliable detection of cherry-picked commits to be out of scope for now, since the returns (in terms of additional included issues) would be small.
    

    We did not tune the parameters of the retrieval methods which support it. However, we used widely recommended defaults \cite{manning_introduction_to_information_retrieval_2008,lv_lower-bounding_2011}.

    \textbf{External Validity}: Our results show strong dataset-specific tendencies. Performance numbers vary by dataset. Moreover, we found that the impact of issue type varies by project and organisation. The same holds for the number of identifiers. Hence, our results provide little guarantees for projects outside of our dataset. 

    Moreover, our results may not generalise well to different types of models. For instance, the information retrieval methods we used only consider word frequencies, and not words in their context like e.g. transformer models.

    In our work, we analysed the results of the method without sub-token splitting, while sub-token splitting achieved comparable performance and better top 5/10 performance. Hence, we also performed our statistical analysis for the method with sub-token splitting (included in our replication package \cite{study-replication-package}). In the end, we found the results to be similar, leading to the same conclusions.

    \textbf{Reliability}: Reliability deals with the replicability of a study. Our dataset was created automatically, and the retrieval methods we investigated are deterministic. The only part of our work involving manual annotation is determining the categories of issue type. To ensure replicability, the categorisation is included in our replication package \cite{study-replication-package}. Moreover, it contains all our code, intermediate results, archive of issue data, and the dataset we created for this  study.
    

\section{Conclusion \& Future Work}\label{sec:conclusion}
    We performed an exploratory analysis of file localisation for general issue types. We provide a dataset and reusable tooling for dataset creation. We investigated the performance of several information retrieval methods, and investigated the performance impacts of issue types and localisation hints in the form of identifiers. Our results show that BM25 is the best performing method, with the bug-specific rVSM performing worst. Both issue type and \# identifiers have project-specific effects with small effect sizes at most. 

    Future work could involve developing more complex models for general file localisation. Examples include experimenting with more context aware text models, and the inclusion of historical information such as co-change. Future work should also keep investigating project and issue-type specific properties of any new features to maintain a good understanding regarding the performance effects, and to develop methods which can be adjusted to project-specific properties. 

\bibliographystyle{ACM-Reference-Format}
\bibliography{references,references-2}

\end{document}